\documentclass[aps,prd,nofootinbib,twocolumn,longbibliography]{revtex4-2}
\usepackage[english]{babel}
\usepackage[CJKbookmarks, pdftex, bookmarksnumbered, bookmarksopen, colorlinks, citecolor=blue, linkcolor=blue]{hyperref}
\usepackage{amsmath,mathrsfs,amssymb,url}
\usepackage{graphicx,xcolor}
\usepackage{enumitem}
\usepackage{cleveref}

\newcommand*{\diff}{\mathop{}\!\mathrm{d}}

\begin{document}

\title{Global Kruskal-Szekeres coordinates for Reissner-Nordström spacetime}

\author{Farshid Soltani}

\affiliation{Department of Physics and Astronomy, University of Western Ontario, London, Ontario N6A 3K7, Canada}

\begin{abstract} 

I derive a smooth and global Kruskal-Szekeres coordinate chart for the maximal extension of the Reissner-Nordström geometry that provides a generalization to the standard inner and outer Kruskal-Szekeres coordinates. The Kruskal-Szekeres diagram associated to this coordinate chart, whose existence is an interesting fact in and of itself, provides a simple alternative to the conformal diagram of the spacetime.

\end{abstract}

\maketitle

\section{Introduction}

The Reissner-Nordström geometry~\cite{Reissner,Nordstrom} describes the gravitational field of an electrically charged spherically-symmetric static black hole. The global causal structure of this spacetime is considerably different from the causal structure of Schwarzschild spacetime, which describes the gravitational field of an electrically neutral spherically-symmetric static black hole, and it is strongly dependent on the relative strength of its gravitational mass $m$ and its electric charge $q$.\footnote{Planck units $c=G=\hbar=1$ will be used throughout the article.} A Reissner-Nordström black hole with $m^2>q^2$ is called non-extremal and, instead of a spacelike curvature singularity hidden behind a single horizon, it features a timelike curvature singularity concealed behind two separate horizons. A Reissner-Nordström black hole with $m^2=q^2$ is called extremal and it features a timelike curvature singularity concealed behind a single horizon. In both cases, given the timelike nature of the singularity, a massive observer falling inside the black hole is not bound to end their wordline at the singularity, and so spacetime must continue in the future of the singularity. In fact, the conformal diagram of the maximal extension of both the extremal and the non-extremal Reissner-Nordström geometry consists of an infinitely periodic tower of asymptotically flat exterior regions connected to each other by black hole interiors.

Focusing for the moment on the non-extremal case, several coordinate charts can be used to cover part or the entirety of the maximal extension of the non-extremal Reissner-Nordström geometry. Some of the most renowned ones, that will also be relevant in the following, are the Eddington-Finkelstein coordinates~\cite{Eddington,Finkelstein:1958zz}, the Kruskal-Szekeres coordinates~\cite{Kruskal:1959vx,Szekeres} and the global coordinate chart of the conformal diagrams defined in~\cite{Carter:1966zza,CarterRNext,Graves:1960zz,Hawking_Ellis,Chandra}. The Eddington-Finkelstein charts use coordinates that are adapted to either ingoing or outgoing null geodesics in order to penetrate the horizons and simultaneously cover one of the infinitely many asymptotically flat exterior regions, the trapped region between the two horizons separating the exterior region from the interior region, and the region inside the inner horizon down to the curvature singularity. They are not global coordinates, but multiple ingoing and outgoing Eddington-Finkelstein charts can be used to tessellate the complete spacetime of the maximal extension of Reissner-Nordström geometry.

The Kruskal-Szekeres charts provide a different set of horizon-penetrating coordinates. One way to derive them, as I will do later on, is by studying the behavior of null geodesics in Eddington-Finkelstein coordinates and to use the results of this analysis to define new coordinates that will be well behaved where the Eddington-Finkelstein coordinates are not. This standard procedure results in the construction of two coordinate charts: The outer and the inner Kruskal-Szekeres charts. The former is able to simultaneously cover one of the infinitely many asymptotically flat exterior regions, the trapped and anti-trapped regions between the two horizons respectively in the future and in the past of this exterior region, and its mirror asymptotic region. The latter is able to simultaneously cover one of the infinitely many interior regions containing a curvature singularity, the anti-trapped and trapped regions between the two horizons respectively in the future and in the past of this interior region, and its mirror interior region. Differently from the Kruskal-Szekeres coordinates in the maximal extension of Schwarzschild spacetime, where they provide a global coordinate chart, the standard Kruskal-Szekeres coordinate charts in non-extremal Reissner-Nordström spacetime are adapted to a specific horizon and fail to cover any region beyond the other horizon.

The coordinates that are used to draw the conformal diagram of the maximal extension of non-extremal Reissner-Nordström geometry provide a global chart for the spacetime. However, differently from the coordinate charts already mentioned, which are all smooth $(C^{\infty})$, the coordinates of the conformal diagram only provide a $C^{2}$ chart.

In this article I derive a smooth Kruskal-Szekeres coordinate chart that covers the complete maximal extension of the non-extremal Reissner-Nordström geometry. This coordinate chart can be considered as a smooth generalization of the $C^{1}$ global Kruskal-Szekeres coordinate chart derived in~\cite{HamiltonGR}. The same construction can be carried out also in the extremal case, thus defining a smooth Kruskal-Szekeres coordinate chart that covers the complete maximal extension of the extremal Reissner-Nordström geometry. The only other smooth (analytic in fact) and global coordinate chart for the maximal extension of both extremal and non-extremal Reissner-Nordström geometry is the (Israel-)Klösch-Strobl coordinate chart\footnote{The Israel coordinate chart firstly introduced in \cite{Israel:1966zz} provides a not widely known analytic and global covering of the maximal extension of the Schwarzschild geometry. The same coordinate chart was later rediscovered by Pajerski and Newman in~\cite{Pajerski:1971tca} and by Klösch and Strobl in~\cite{Klosch:1995bw}. In~\cite{Israel:1966zz} Israel also provides a generalization of this chart to the non-extremal Reissner-Nordström geometry. However, as explained in~\cite{Klosch:1995bw}, this coordinate chart does not provide a global covering of the maximal extension of non-extremal Reissner-Nordström geometry.}~\cite{Israel:1966zz,Klosch:1995bw}.

The global Kruskal-Szekeres coordinate chart I derive here is not specific to the Reissner-Nordström spacetime but it can be straightforwardly generalized to any static spherically-symmetric black hole spacetime whose geometry include multiple horizons. It has indeed been recently used in~\cite{PG_OS} to draw the global Kruskal-Szekeres diagram of the geometry of a quantum modification of the Oppenheimer-Snyder model.

The article is structured in the following way. In \cref{secRNgeometry} I briefly review the non-extremal Reissner-Nordström geometry and its causal structure, and I introduce the ingoing and outgoing Eddington-Finkelstein coordinate charts. Inner and outer Kruskal-Szekeres coordinate charts are derived in \cref{secinoutKS}. The smooth and global Kruskal-Szekeres coordinate charts for the non-extremal and the extremal Reissner-Nordström geometry are derived respectively in \cref{secglobalKS} and in \cref{secxtrm}. Finally, I make a few closing remarks in \cref{secconclusions}.

\section{The Reissner-Nordström geometry}
\label{secRNgeometry}

The Reissner-Nordström metric reads
\begin{equation}
\diff s^2 = - h(r)\diff t^2 + h^{-1}(r)  \diff r^2 + r^2 \diff \Omega^2\, ,
\end{equation}
\begin{equation}
h(r)=1-\frac{2m}{r}+\frac{q^2}{r^2}\,,
\end{equation}
where $m$ and $q$ are the mass and the electric charge of the black hole, $\diff \Omega^2=\diff \theta ^2 + \sin^2 (\theta) \diff \phi^2$ is the metric of a unit two-sphere and $(t,r,\theta,\phi)$ is the spherical coordinate system adapted to static observers moving along the asymptotically timelike Killing vector field $\partial_t$. I will focus on the non-extremal case of $m^2>q^2$ up to \cref{secxtrm}, where instead the extremal case will be discussed. 

\begin{figure}[b]
	\centering
\includegraphics[scale=.65]{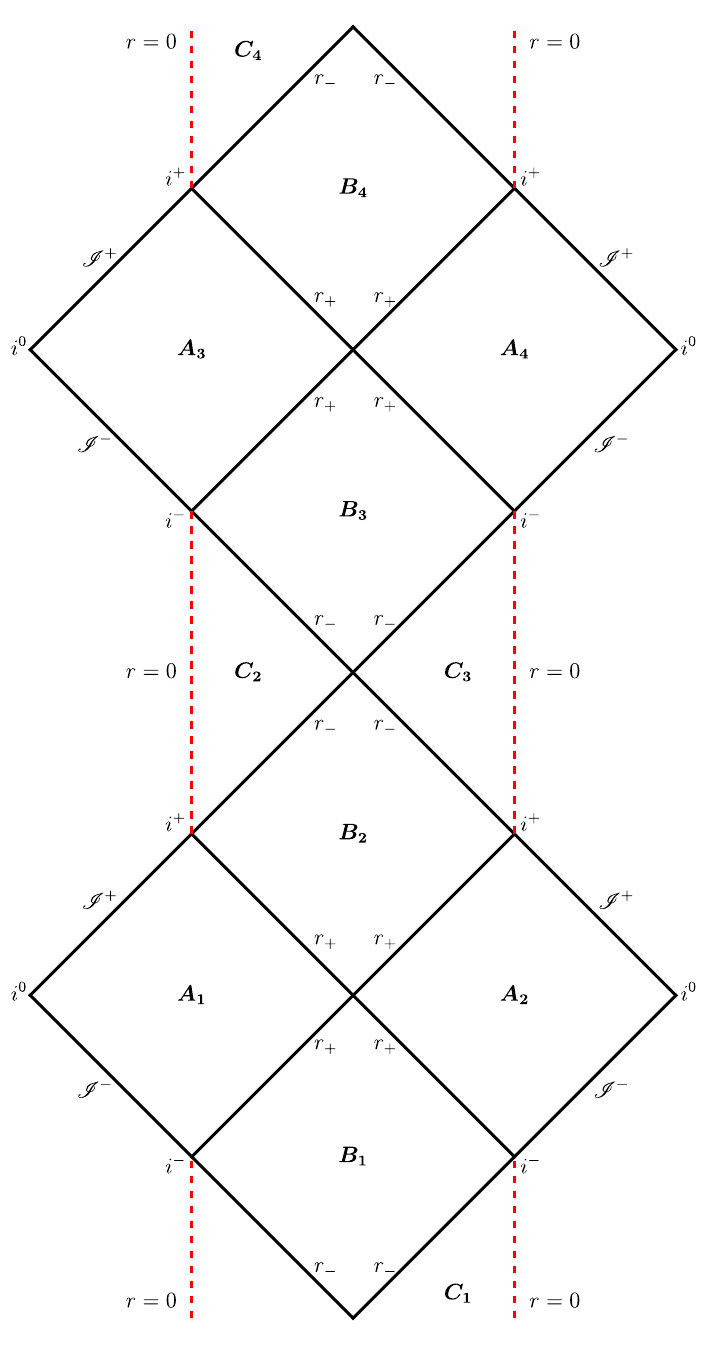}
\caption{Conformal diagram of the maximal extension of non-extremal Reissner-Nordström spacetime.}
\label{figRNconformal}
\end{figure}

The maximal extension of this geometry, whose conformal diagram is reported in \cref{figRNconformal}, is well known and studied in the literature. For $m^2>q^2$, the function $h(r)$ has two zeros $r_\pm=m\pm \sqrt{m^2-q^2}$, which result in two different horizons of the black hole: an outer horizon at $r=r_+$ and an inner horizon at $r=r_-$. The killing vector field $\partial_t$ is timelike for $r>r_+$ and $r<r_-$, spacelike for $r_-<r<r_+$ and null for $r=r_\pm$. Inside the inner horizon there is a timelike curvature singularity at $r=0$. Differently from Schwarzschild spacetime, an observer that falls inside the outer horizon is not bound to hit the curvature singularity but can instead exit the black hole interior into a second asymptotic region in the future of the first one. In fact, there is an infinite tower of asymptotically flat exterior regions connected to each other by black hole interiors. The spacetime can then be subdivided in different regions separated by the horizons as shown in \cref{figRNconformal}: Asymptotic regions $A_i$ where $r>r_+$, innermost regions $C_i$ where $r<r_-$, and trapped or anti-trapped regions $B_i$ where $r_-<r<r_+$. The index $i$ takes values in $\mathbb{Z}$.

The static coordinate system $(t,r,\theta,\phi)$ can separately cover each one of the infinitely many regions $A_i$, $B_i$ and $C_i$, with the time coordinate $t$ going to infinity on every horizon. Horizon-penetrating coordinates can be constructed introducing the tortoise coordinate $r^*$ satisfying
\begin{equation}
    \diff r^* = \frac{\diff r}{h(r)}\,.
    \label{tortuga}
\end{equation}
The solution $r^*(r)$ of this equation is
\begin{equation}
    r^* = r + \frac{1}{2 \kappa_+} \log \Big| \frac{r}{r_+}-1 \Big|  + \frac{1}{2 \kappa_-} \log\Big|\frac{r}{r_-}-1 \Big| + C\,,
    \label{r*}
\end{equation}
where $C$ is a constant of integration and 
\begin{equation}
     \kappa_\pm = \pm \frac{r_+ - r_-}{2 r^2_\pm}
\end{equation}
is the surface gravity associated to the horizon $r_\pm$. It is then possible to introduce the retarded and advanced time null coordinates $u$ and $v$ by
\begin{equation}
    u = t-r^* , \quad \quad  v = t+r^* .
    \label{uv}
\end{equation}
The coordinate chart $(v,r,\theta,\phi)$ is known as ingoing Eddington-Finkelstein chart and the Reissner-Nordström metric in these coordinates reads
\begin{equation}
\diff s^2 = - h(r)\diff v^2 + 2 \diff v \diff r + r^2 \diff \Omega^2\, .
\label{EFin}
\end{equation}
The advanced time coordinate $v$ is well behaved on all the horizons that are represented as $45^\circ$ lines in the diagram in \cref{figRNconformal}, but it goes to infinity on the others. The ingoing Eddington-Finkelstein coordinate system is then able to separately cover all the spacetime regions between two consecutive $-45^\circ$ lines in \cref{figRNconformal}. Namely, it can be used to simultaneously cover regions $A_1$, $B_1$ and $C_1$, or regions $A_2$, $B_2$ and $C_2$, etc.

Analogously, the retarded time coordinate $u$ is well-behaved on all $-45^\circ$ line horizons and it goes to infinity on the others. Thus, the outgoing Eddington-Finkelstein coordinates $(u,r,\theta,\phi)$ separately covers all the spacetime regions between two consecutive $45^\circ$ lines in \cref{figRNconformal}. E.g., it can be used to simultaneously cover regions $A_1$, $B_2$ and $C_3$, or regions $A_4$, $B_3$ and $C_2$, etc.

Interestingly, although the coordinates $u$ and $v$ are separately well defined on several horizons, one or the other is ill defined on every given horizon, and thus the null coordinate system $(u,v,\theta,\phi)$ does not improve the spacetime coverage of the static coordinate system $(t,r,\theta,\phi)$. The Reissner-Nordström metric in the double-null coordinate system $(u,v,\theta,\phi)$ reads
\begin{equation}
\diff s^2 = - h(r)\diff u \diff v + r^2 \diff \Omega^2\,,
\end{equation}
where $r=r(u,v)$ is implicitly defined by
\begin{equation}
r^* (r)= \frac{v-u}{2}\,.
\label{ruvimp}
\end{equation}

\section{Inner and outer Kruskal-Szekeres coordinates}
\label{secinoutKS}

A different set of horizon-penetrating coordinates is the Kruskal-Szekeres coordinates. One way to derive them, that will prove useful also for the global Kruskal-Szekeres chart, is to study the behavior of null geodesics in Eddington-Finkelstein coordinates.

Consider then the spacetime regions $A_2$, $B_2$ and $C_2$ in \cref{figRNconformal} in ingoing Eddington-Finkelstein coordinates. Given the metric in \cref{EFin}, radial null geodesics must satisfy the equations
\begin{equation}
h(r) \dot{v}^2 -2\dot{v}\dot{r}=0
\label{EFeq1}
\end{equation}
(normalization of 4-velocity) and
\begin{equation}
h(r) \dot{v} -\dot{r}=E
\label{EFeq2}
\end{equation}
(due to $v$-translation invariance), where the overdot means differentiation with respect to an affine parameter $\lambda$ and $E$ is the constant of motion associated with the $v$-translation invariance. A first set of solutions is given by geodesics satisfying
\begin{equation}
\dot{v}=0\,,\quad\quad\quad\quad \dot{r}=-E\,,
\end{equation}
that is
\begin{equation}
v(\lambda)=v_0\,,\quad\quad r(\lambda)=r_0-E\lambda\,.
\end{equation}
For $E>0$ these solutions, valid for $\lambda\in(-\infty , r_0/E)$, describe future-oriented ingoing radial null geodesics starting out at past null infinity in $A_2$, diagonally crossing regions $A_2$, $B_2$, $C_2$ and then finally hitting the $r=0$ curvature singularity at $\lambda= r_0/E$.

A different set of solutions, which by exclusion will describe outgoing radial null geodesics, satisfy
\begin{equation}
\dot{v}=\frac{2E}{h(r)}\,,\quad\quad\quad\quad \dot{r}=E\,.
\end{equation}
If we are interested in future-oriented geodesics, $E$ must be taken negative in $B_2$ and positive in $A_2$ and $C_2$. Focusing our attention on region $B_2$, and taking $E=-1$ and $r_0=0$ for simplicity, outgoing radial null geodesics are given by
\begin{equation}
r(\lambda)=-\lambda\,,\quad\quad \lambda\in (-r_+ , -r_-)\,,
\end{equation}
\begin{equation}
v(\lambda)= -2\lambda + \frac{1}{ \kappa_+} \log \Big| \frac{\lambda}{r_+}+1 \Big|  + \frac{1}{ \kappa_-} \log\Big|\frac{\lambda}{r_-}+1 \Big| + K\,,
\label{vout}
\end{equation}
where $K$ is a constant identifying different geodesics. These curves start at the past outer horizon between regions $B_2$ and $A_1$, that is $r\rightarrow r_+$ and $v\rightarrow -\infty$ for $\lambda \rightarrow (-r_+)^{+}$, cross diagonally region $B_2$ and end at the future inner horizon between regions $B_2$ and $C_3$, that is $r\rightarrow r_-$ and $v\rightarrow +\infty$ (notice that $\kappa_-<0$) for $\lambda \rightarrow (-r_-)^{-}$. Naturally, these geodesics do not end at the two horizons, but the null coordinate $v$ is not able to follow them past the horizons.

This analysis perfectly displays the physics of ingoing Eddington-Finkelstein coordinates. They are able to cover the full range of the affine parameter of ingoing radial null geodesics, as they are adapted to them, but they cover only a finite interval $\lambda\in(-r_+ , -r_-)$ of the affine parameter of outgoing radial null geodesics in the full coordinate interval $v\in(-\infty ,+\infty)$. The analysis itself however suggests a solution to this issue: the affine parameter $\lambda$ is the natural coordinate to extend $v$ beyond the horizons.

The same analysis performed in outgoing Eddington-Finkelstein coordinates covering regions $A_1$, $B_2$ and $C_3$ shows that future-oriented outgoing radial null geodesics are given by 
\begin{equation}
u(\sigma)=u_0\,,\quad\quad r(\sigma)=r_0+E\sigma\,,
\end{equation}
for affine parameter $\sigma\in(-\infty,-r_0/E)$ and $E<0$, while future-oriented ingoing radial null geodesics ($E=1$ and $r_0=0$) in region $B_2$ are given by
\begin{equation}
r(\sigma)=-\sigma\,,\quad\quad \sigma\in (-r_+ , -r_-)\,,
\end{equation}
\begin{equation}
u(\sigma)= 2\sigma - \frac{1}{ \kappa_+} \log \Big| \frac{\sigma}{r_+}+1 \Big|  - \frac{1}{ \kappa_-} \log\Big|\frac{\sigma}{r_-}+1 \Big| + K'\,,
\label{uin}
\end{equation}
where $K'$ is a constant identifying different geodesics. Thus, only a finite interval $\sigma\in(-r_+ , -r_-)$ of the affine parameter of ingoing radial null geodesics in region $B_2$ is covered in the full coordinate interval $u\in(-\infty ,+\infty)$ and the affine parameter $\sigma$ is the natural coordinate to extend $u$ beyond the horizons.

So, starting from region $B_2$ in the null coordinate system $(u,v,\theta,\phi)$, in order to obtain coordinates that are well defined on both the outer horizons, the discussion above suggests to make the following change of coordinates (keeping only the leading term near the outer horizon in \cref{vout,uin}):
\begin{equation}
u(U_+)= - \frac{1}{ \kappa_+} \log | U_+|\, ,
\label{uU+}
\end{equation}
\begin{equation}
v(V_+)=  \frac{1}{ \kappa_+} \log | V_+ |\, ,
\label{vV+}
\end{equation}
where $U_+$ and $V_+$ are adimensional null coordinates such that $U_+>0$ and $V_+>0$ in $B_2$ and the position of the outer horizons has been set to $U_+=0$ and $V_+=0$. The non-extremal Reissner-Nordström metric in $(U_+,V_+,\theta,\phi)$ coordinates, also known as outer Kruskal-Szekeres null coordinates, reads
\begin{equation}
\diff s^2 =  \frac{1}{\kappa_+^2}\frac{h(r)}{U_+ V_+} \diff U_+ \diff V_+ + r^2 \diff \Omega^2\,,
\end{equation}
where $r=r(U_+,V_+)$ is implicitly defined by \cref{ruvimp,uU+,vV+}. It is straightforward to check that the metric tensor is indeed well-defined on all outer horizons and that the outer Kruskal-Szekeres null coordinates simultaneously cover regions $A_1$, $B_1$, $A_2$ and $B_2$, going to infinity on all inner horizons. These coordinates can in fact be used to separately cover all the blocks of spacetime regions divided by outer horizons such as $A_3$, $B_3$, $A_4$ and $B_4$, etc. 

If we take the outer Kruskal-Szekeres null coordinates to simultaneously cover regions $A_1$, $B_1$, $A_2$ and $B_2$, then the transformation in \cref{uU+,vV+} separately gives a well-defined diffeomorphism between $(U_+,V_+,\theta,\phi)$ and $(u,v,\theta,\phi)$ in each of the regions $A_1$, $B_1$, $A_2$ and $B_2$. Being the null coordinate system $(u,v,\theta,\phi)$ singular on the horizons, also the transformation in \cref{uU+,vV+} is ill defined there.

Analogously, in order to obtain coordinates that are well defined on both the inner horizons, inner Kruskal-Szekeres null coordinates $(U_-, V_-)$ should be defined as
\begin{equation}
u(U_-)= - \frac{1}{ \kappa_-} \log | U_--1|\, ,
\label{uU-}
\end{equation}
\begin{equation}
v(V_-)=  \frac{1}{ \kappa_-} \log | V_- -1|\, ,
\label{vV-}
\end{equation}
where $U_-<1$ and $V_-<1$ in $B_2$ and the position of the inner horizons has been set to $U_-=1$ and $V_-=1$. The non-extremal Reissner-Nordström metric in inner Kruskal-Szekeres coordinates $(U_-,V_-,\theta,\phi)$ reads
\begin{equation}
\diff s^2 =  \frac{1}{\kappa_-^2}\frac{h(r)}{(U_- -1) (V_- -1)} \diff U_- \diff V_- + r^2 \diff \Omega^2\,,
\end{equation}
where $r=r(U_-,V_-)$ is implicitly defined by \cref{ruvimp,uU-,vV-}. This metric tensor is well-defined on all inner horizons and the inner Kruskal-Szekeres coordinates simultaneously cover regions $B_2$, $C_2$, $B_3$ and $C_3$ (or any block of spacetime regions divided by inner horizons), going to infinity on all outer horizons.

If we take the inner Kruskal-Szekeres null coordinates to simultaneously cover regions $B_2$, $C_2$, $B_3$ and $C_3$, then the transformation in \cref{uU-,vV-} separately gives a well-defined diffeomorphism between $(U_-,V_-,\theta,\phi)$ and $(u,v,\theta,\phi)$ in each of the regions $B_2$, $C_2$, $B_3$ and $C_3$. Being the null coordinate system $(u,v,\theta,\phi)$ singular on the horizons, also the transformation in \cref{uU-,vV-} is ill defined there.

While very interesting coordinate systems, these Kruskal-Szekeres coordinate charts for the non-extremal Reissner-Nordström metric are not as convenient as the Kruskal-Szekeres coordinates for the Schwarzschild metric, as the latters provide a global covering of spacetime. In the next section I generalize the construction given in this section in order to define a Kruskal-Szekeres coordinate chart covering the complete maximal extension of the non-extremal Reissner-Nordström geometry.

\begin{figure}[t]
	\centering
\includegraphics[scale=.7]{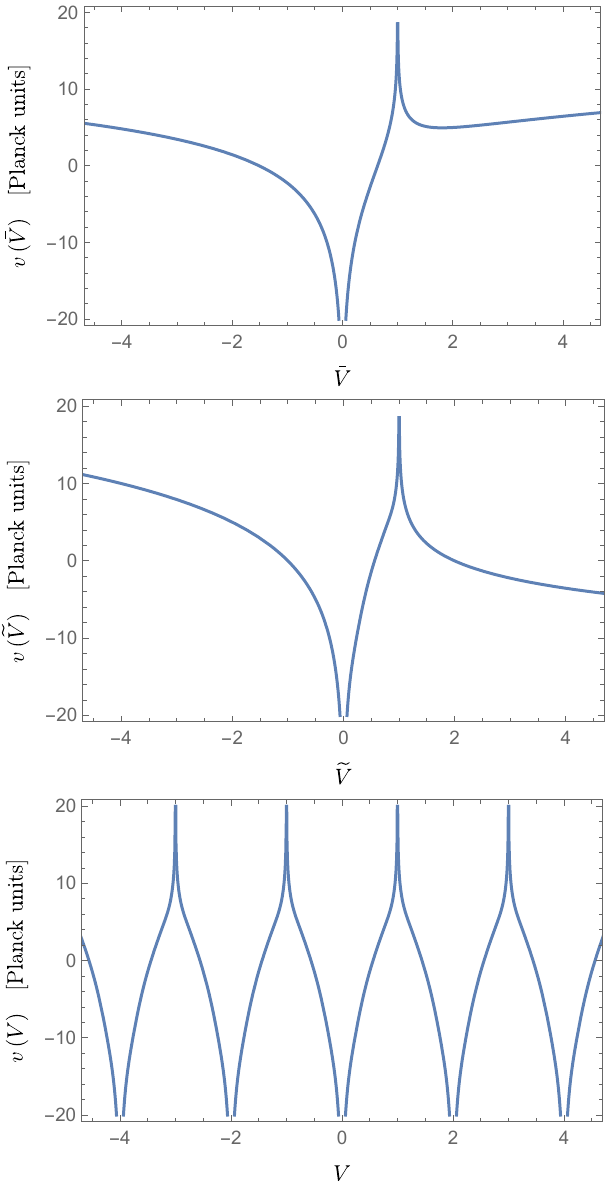}
\caption{Plot of $v(\bar{V})$ in \cref{vVbar} (top), $v(\widetilde{V})$ in \cref{vVtilde} (center), and $v(V)$ in \cref{vV} (bottom) with $m=1$ and $q=0.98$ (Planck units).} 
\label{figVbarVtilde}
\end{figure}

\section{Global Kruskal-Szekeres chart}
\label{secglobalKS}

Let me start with finding a Kruskal-Szekeres coordinate system covering region $B_2$ that is well behaved on both inner and outer horizons. From the discussion in the previous section, the natural choice for such coordinates would be
\begin{equation}
u(\bar{U})=- \frac{1}{ \kappa_+} \log | \bar{U}| - \frac{1}{ \kappa_-} \log | \bar{U} -1|\, ,
\label{uUbar}
\end{equation}
\begin{equation}
v(\bar{V})= \frac{1}{ \kappa_+} \log | \bar{V}| + \frac{1}{ \kappa_-} \log | \bar{V} -1|,
\label{vVbar}
\end{equation}
where $\bar{U}$ and $\bar{V}$ are adimensional null coordinates such that the outer horizons have been set to $\bar{U}=0$, $\bar{V}=0$ and the inner horizons have been set to $\bar{U}=1$, $\bar{V}=1$. It is however straightforward to see that this is not a well-defined change of coordinate, as the function $v(\bar{V})$ ($u(\bar{U})$) is not separately monotonic in each of the regions 
\begin{align*}
\begin{split}
\bar{V}<0\quad&\quad(\bar{U}<0)\,,\\
0<\bar{V}<1\quad&\quad(0<\bar{U}<1)\,,\\
\bar{V}>1\quad&\quad(\bar{U}>1)\,.
\end{split}
\end{align*}

Focusing on the behavior of $v(\bar{V})$ shown in the top panel of \cref{figVbarVtilde} for a specific choice of $m$ and $q$, the problem is that the logarithm associated to the outer horizon at $\bar{V}=0$ dominates the asymptotic behavior of $v(\bar{V})$ in both $\bar{V}<0$ and $\bar{V}>1$, while it should only dominate in $\bar{V}<0$, with the other logarithm dominating the asymptotic behavior in $\bar{V}>1$. This issue can be taken care of with a smooth transition function that interpolates between $0$ and $1$ in such a way that, when multiplied to the logarithms, it ensures that their contribution is only restricted to the desired regions.

Consider in fact the function
\begin{equation}
g_{\uparrow}(X)=\frac{F(X)}{F(X) + F(1-X)}\, ,
\label{gup}
\end{equation}
with 
\begin{equation}
F(X)= 
 \begin{cases}
     0 &\quad X\leq 0 \, ,\\ 
       e^{-1/X} &\quad X>0\, .\\
     \end{cases}
\end{equation}
The function $g_{\uparrow}(X)$ is smooth, it satisfies $g_{\uparrow}(X)=0$ for $X\leq0$ and $g_{\uparrow}(X)=1$ for $X \geq 1$, and its derivatives of all orders in $X=0$ and $X=1$ are vanishing. It is thus a function that smoothly interpolates between $0$ and $1$ in the interval $X\in[0,1]$. A function that smoothly interpolates between $1$ and $0$ in the interval $X\in[0,1]$ is given by 
\begin{equation}
g_{\downarrow}(X)=1-g_{\uparrow}(X)\, .
\label{gdown}
\end{equation}

\begin{figure}[t]
	\centering
\includegraphics[scale=.6]{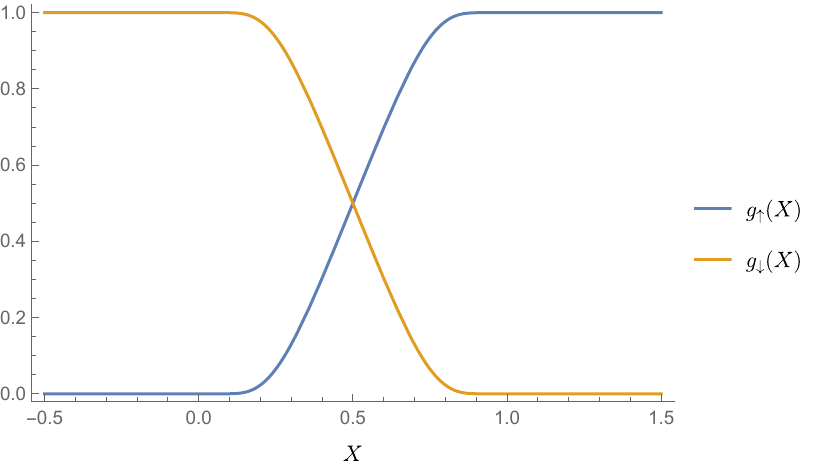}
\caption{Plot of the functions $g_{\uparrow}(X)$ and $g_{\downarrow}(X)$ defined in \cref{gup,gdown}.} 
\label{figgupdown}
\end{figure}

These functions, whose behavior is plotted in \cref{figgupdown}, can be used to modify the change of coordinate in \cref{uUbar,vVbar} in the following way:
\begin{equation}
u(\widetilde{U})=- \frac{1}{ \kappa_+} g_{\downarrow}(\widetilde{U}) \log | \widetilde{U}| - \frac{1}{ \kappa_-} g_{\uparrow}(\widetilde{U}) \log | \widetilde{U} -1|\, ,
\label{uUtilde}
\end{equation}
\begin{equation}
v(\widetilde{V})= \frac{1}{ \kappa_+} g_{\downarrow}(\widetilde{V}) \log | \widetilde{V}| + \frac{1}{ \kappa_-} g_{\uparrow}(\widetilde{V})\log | \widetilde{V} -1| .
\label{vVtilde}
\end{equation}
The function $v(\widetilde{V})$ ($u(\widetilde{U})$) is now separately monotonic in each of the regions 
\begin{align*}
\begin{split}
\widetilde{V}<0\quad&\quad(\widetilde{U}<0)\,,\\
0<\widetilde{V}<1\quad&\quad(0<\widetilde{U}<1)\,,\\
\widetilde{V}>1\quad&\quad(\widetilde{U}>1)\,,
\end{split}
\end{align*}
as it can be checked from the plot of $v(\widetilde{V})$ in the center panel of \cref{figVbarVtilde}. The non-extremal Reissner-Nordström metric in these double-horizon-penetrating Kruskal-Szekeres coordinates $(\widetilde{U},\widetilde{V},\theta,\phi)$ reads
\begin{equation}
\diff s^2 = h(r) \widetilde{f}(\widetilde{U}) \widetilde{f}(\widetilde{V}) \diff  \widetilde{U} \diff \widetilde{V} + r^2 \diff \Omega^2\,,
\label{metrictilde}
\end{equation}
where $r=r(\widetilde{U},\widetilde{V})$ is implicitly defined by \cref{ruvimp,uUtilde,vVtilde} and
\begin{equation}
\begin{split}
\widetilde{f}(X)=& \frac{1}{ \kappa_+} g'_{\downarrow}(X) \log | X| + \frac{1}{ \kappa_+} \frac{g_{\downarrow}(X)}{X} \\
& +
\frac{1}{ \kappa_-} g'_{\uparrow}(X) \log | X-1| + \frac{1}{ \kappa_-} \frac{g_{\uparrow}(X)}{X-1} \,.
\label{ftilde}
\end{split}
\end{equation}
Exactly as in the case of outer and inner Kruskal-Szekeres coordinates, it is easy to check that the metric tensor in \cref{metrictilde} is indeed well-defined on all outer and inner horizons bounding region $B_2$, and that this Kruskal-Szekeres null coordinate chart simultaneously cover regions $A_1$, $B_1$, $A_2$, $B_2$, $C_2$, $B_3$, and $C_3$, going to infinity on the inner horizons bounding $B_1$ and the outer horizons bounding $B_3$. They can in fact be used to separately cover all the blocks of spacetime regions such as the one just described.

The transformation in \cref{uUtilde,vVtilde} separately gives a well-defined diffeomorphism between $(\widetilde{U},\widetilde{V},\theta,\phi)$ and $(u,v,\theta,\phi)$ in each spacetime region covered by the latter, even if an analytical expression for the inverse transformation cannot be given. In fact, restricting our attention to one of these regions, e.g. $B_2$ $(0<\widetilde{U}<1,0<\widetilde{V}<1)$, the strictly monotonicity of $u(\widetilde{U})$ and $v(\widetilde{V})$ in this region ensures the invertibility of the transformation and the inverse function theorem ensures the differentiability of its inverse.\footnote{In the general case the statement of the inverse function theorem holds true only locally. However, for real functions of one real variable the statement holds true globally.} Being the null coordinate system $(u,v,\theta,\phi)$ singular on the horizons, also the transformation in \cref{uUtilde,vVtilde} is ill defined there.

We thus succeeded in constructing a Kruskal-Szekeres coordinate chart able to simultaneously cover two successive outer and inner horizons. A Kruskal-Szekeres coordinate chart able to simultaneously cover two successive inner and outer horizons can be constructed analogously. However, not much would be gained by this. What we want to find now is a global Kruskal-Szekeres coordinate chart covering the entirety of the maximal extension of non-extremal Reissner-Nordström spacetime with its infinite tower of asymptotic regions and black hole interiors.

The way to accomplish this is to use the same construction used to cover two successive horizons multiple times, in such a way to cover the infinite tower of horizons. Consider in fact the following change of coordinates:
\begin{widetext}
\begin{equation}
u(U)=
 \begin{cases}
    \quad\quad\quad : &\\
    - \frac{1}{ \kappa_+} g_{\uparrow}(U,2n-1,2n) \log | U-2n| - \frac{1}{ \kappa_-} g_{\downarrow}(U,2n-1,2n) \log | U - (2n-1)| &\quad 2n-1<U  <2n \, ,\\ 
    - \frac{1}{ \kappa_+} g_{\downarrow}(U,2n,2n+1) \log | U-2n| - \frac{1}{ \kappa_-} g_{\uparrow}(U,2n,2n+1) \log | U - (2n+1)| &\quad 2n<U  <2n+1 \, ,\\ 
    \quad\quad\quad : &\\
\end{cases}
\label{uU}
\end{equation}
\begin{equation}
v(V)=
 \begin{cases}
    \quad\quad\quad : &\\
    \frac{1}{ \kappa_+} g_{\uparrow}(V,2n-1,2n) \log | V-2n| + \frac{1}{ \kappa_-} g_{\downarrow}(V,2n-1,2n) \log | V- (2n-1)| &\quad 2n-1<V  <2n \,, \\ 
    \frac{1}{ \kappa_+} g_{\downarrow}(V,2n,2n+1) \log | V-2n| + \frac{1}{ \kappa_-} g_{\uparrow}(V,2n,2n+1) \log | V - (2n+1)| &\quad 2n<V <2n+1 \, ,\\ 
    \quad\quad\quad : &\\
\end{cases}
\label{vV}
\end{equation}
\end{widetext}
with $n\in\mathbb{Z}$ and $g_{\uparrow\downarrow} (X,a,b)=g_{\uparrow\downarrow} \big(\frac{X-a}{b-a}\big)$. The function $v(V)$ is plotted in the bottom panel of \cref{figVbarVtilde}. The position of the horizons in the $(U,V)$ coordinates is arbitrary. I made the choice $U=n$ and $V=n$ for $n\in\mathbb{Z}$. 

\begin{figure}[ht]
	\centering
\includegraphics[scale=.6]{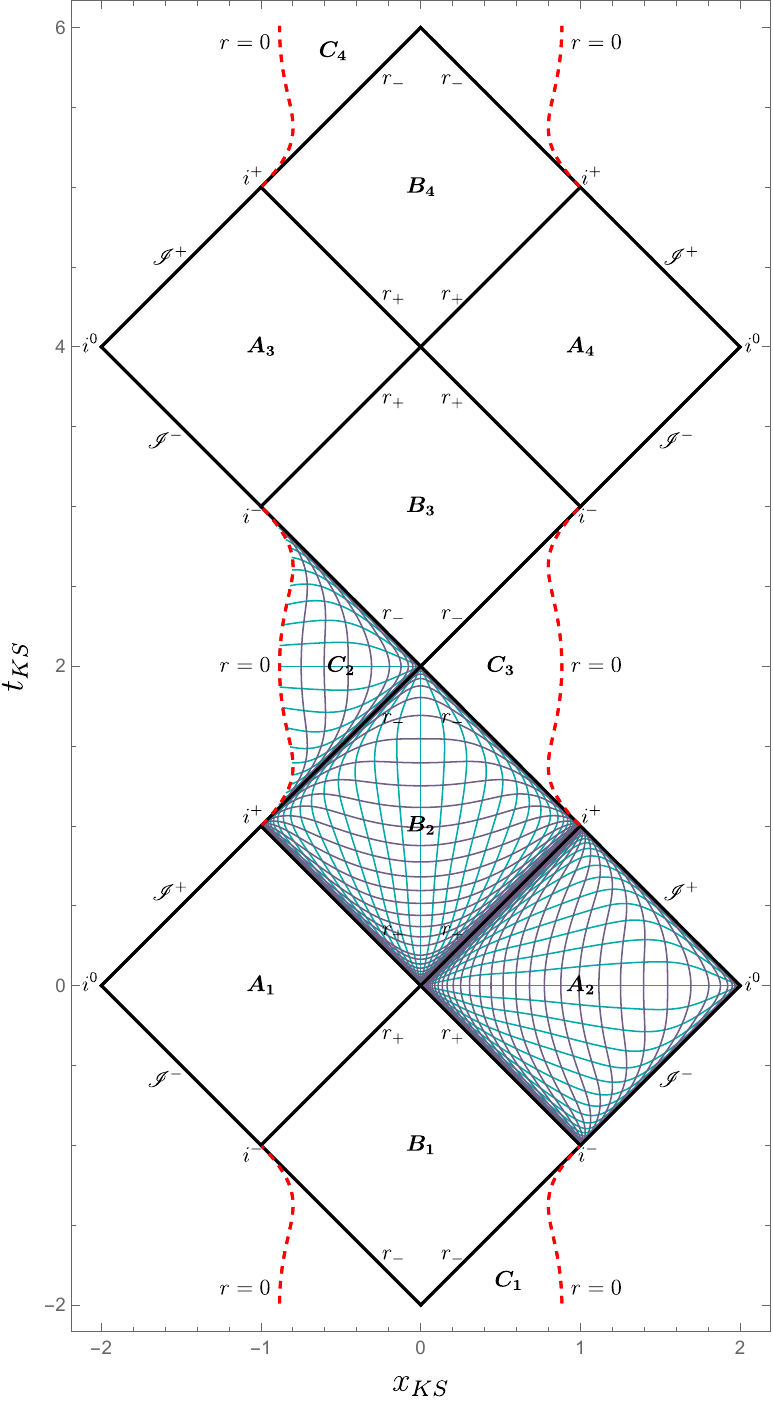}
\caption{Kruskal-Szekeres diagram of the maximal extension of non-extremal Reissner-Nordström geometry with $m=1$ and $q=0.98$ (Planck units). In light blue curves of constant $t$ and in violet curves of constant $r$.}
\label{figKSdiagram}
\end{figure}


Defining the function $f(X)$ in such a way that $v'(V)=f(V)$ and $u'(U)=-f(U)$, the non-extremal Reissner-Nordström metric in the Kruskal-Szekeres coordinates $(U,V,\theta,\phi)$ reads
\begin{equation}
\diff s^2 = h(r) f(U)f(V) \diff U \diff V + r^2 \diff \Omega^2\,,
\label{metricUV}
\end{equation}
where $r=r(U,V)$ is implicitly defined by \cref{ruvimp,uU,vV}. These equations can then be used to study the regularity of both the function $r(U,V)$ itself and of the metric tensor near the horizons. The analysis is carried out exactly as it is carried out for the standard inner and outer Kruskal-Szekeres null coordinates, and it shows that the metric tensor is well behaved on all spacetime horizons and that the Kruskal-Szekeres null coordinates $(U,V,\theta,\phi)$ provide a smooth and global chart for the maximal extension of the non-extremal Reissner-Nordström geometry.

The transformation in \cref{uU,vV} separately gives a well-defined diffeomorphism between $(U,V,\theta,\phi)$ and $(u,v,\theta,\phi)$ in each spacetime region covered by the latter, even if an analytical expression for the inverse transformation cannot be given. Once again, this is ensured by the strictly monotonicity of $u(U)$ and $v(V)$ (see \cref{figVbarVtilde}) in each spacetime region covered by the $(u,v,\theta,\phi)$ coordinate system and by the inverse function theorem. Being the double-null coordinates $(u,v,\theta,\phi)$ singular on the horizons, also the transformation in \cref{uU,vV} is ill defined there.

Kruskal-Szekeres spatial $x_{KS}$ and temporal $t_{KS}$ coordinates can be defined as
\begin{equation}
t_{KS}= V+U\,,\quad\quad x_{KS}=V-U\,.
\label{xtKS}
\end{equation}
The Kruskal-Szekeres diagram of the maximal extension of non-extremal Reissner-Nordström spacetime is reported in \cref{figKSdiagram}. This diagram is very similar to the conformal diagram in \cref{figRNconformal}. The main reason for this is that future and past null infinity of the asymptotic regions are located at the same Eddington-Finkelstein coordinate value, either $v\rightarrow \pm \infty$ or $u\rightarrow \pm \infty$, of the inner horizons. So, since the inner horizons are at a finite coordinate value in the global Kruskal-Szekeres coordinates, so are future and past null infinity of the asymptotic regions, as in a conformal diagram.

Equivalently, the same result can also be obtained with the more compact change of coordinates
\begin{equation}
\begin{split}
u(U)= -\sum_{n\in\mathbb{Z}} & \Big[ B(U,2n) \frac{\log |U-2n|}{\kappa_+} \\
& + B(U,2n+1) \frac{\log |U-(2n+1)|}{\kappa_-} \Big]\,,
\label{uUb}
\end{split}
\end{equation}
\begin{equation}
\begin{split}
v(V)=\sum_{n\in\mathbb{Z}} & \Big[ B(V,2n) \frac{\log |V-2n|}{\kappa_+} \\
& + B(V,2n+1) \frac{\log |V-(2n+1)|}{\kappa_-} \Big]\, ,
\label{vVb}
\end{split}
\end{equation}
where $B(X,a)$ is the bump function
\[
B(X,a)=
 \begin{cases}
      \exp \Big(- \frac{1}{1-(X-a)^2} \Big) &\quad a-1<X <a+1 \, , \\ 
       0 &\quad \text{otherwise} \,. \\ 
     \end{cases}
\label{bump}
\]

\section{Extremal case}
\label{secxtrm}

The exact same construction can be carried out also in the extremal case $m^2=q^2$, resulting in a global Kruskal-Szekeres coordinate chart covering the entirety of the maximal extension of the extremal Reissner-Nordström spacetime.

\begin{figure}[t]
	\centering
\includegraphics[scale=.9]{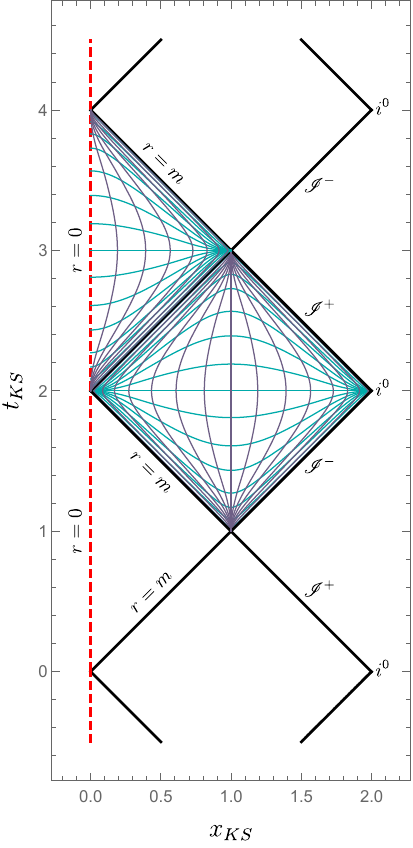}
\caption{Kruskal-Szekeres diagram of the maximal extension of the extremal Reissner-Nordström geometry with $m=1$ (Planck units). In light blue curves of constant $t$ and in violet curves of constant $r$.}
\label{figKSdiagramx}
\end{figure}

The extremal Reissner-Nordström metric in the static coordinate system $(t,r,\theta,\phi)$ still reads
\begin{equation}
\diff s^2 = - h(r)\diff t^2 + h^{-1}(r)  \diff r^2 + r^2 \diff \Omega^2\, ,
\end{equation}
but now
\begin{equation}
h(r)=\Big(1-\frac{m}{r}\Big)^2\,.
\label{hx}
\end{equation}
This function has only a single (double) zero at $r=m$, which means that an extremal Reissner-Nordström black hole only have a single horizon. The maximal extension of this geometry is shown in \cref{figKSdiagramx}.

The tortoise coordinate $r^*$ satisfying \cref{tortuga} is now given by
\begin{equation}
    r^* = r + 2m \log \Big| \frac{r}{m}-1 \Big|  -\dfrac{m}{\frac{r}{m}-1} + C'\,,
    \label{r*x}
\end{equation}
where $C'$ is a constant of integration. Retarded and advanced time null coordinates $u$ and $v$ are defined as in \cref{uv}. Ingoing $(v,r,\theta,\phi)$ and outgoing $(u,r,\theta,\phi)$ Eddington-Finkelstein coordinates and double-null coordinates $(u,v,\theta,\phi)$ can then be used to cover different regions of the maximal extension of the extremal Reissner-Nordström spacetime.

The analysis of the behavior of null geodesics in Eddington-Finkelstein coordinates for the extremal geometry closely follow the analysis of the non-extremal case discussed in \cref{secinoutKS}. Outgoing radial null geodesics in ingoing Eddington-Finkelstein coordinates satisfy
\begin{equation}
\dot{v}=\frac{2E}{h(r)}\,,\quad\quad\quad\quad \dot{r}=E\,,
\end{equation}
where $h(r)$ is now given by \cref{hx}. Focusing our attention to one of the asymptotically flat exterior regions, and taking $r_0=0$ and $E=1$ (future-oriented geodesics have $E>0$) for simplicity, outgoing radial null geodesics are given by
\begin{equation}
r(\lambda)=\lambda\,,\quad\quad \lambda\in (m , \infty)\,,
\end{equation}
\begin{equation}
v(\lambda)= 2\lambda + 4m \log \Big| \frac{\lambda}{m}-1 \Big|  -\dfrac{2m}{\frac{\lambda}{m}-1} + K\,,
\label{voutx}
\end{equation}
where $K$ is a constant identifying different geodesics. These curves start at the past horizon of the exterior region, that is $r\rightarrow m$ and $v\rightarrow -\infty$ for $\lambda \rightarrow m^{+}$, cross diagonally this exterior region and end at future null infinity $\mathscr{I}^+$.

Similarly, future-oriented ingoing radial null geodesics ($E=1$ and $r_0=0$) in outgoing Eddington-Finkelstein coordinates are given by
\begin{equation}
r(\sigma)=-\sigma\,,\quad\quad \sigma\in (-\infty , -m)\,,
\end{equation}
\begin{equation}
u(\sigma)= 2\sigma -  4m \log \Big| \frac{\sigma}{m}+1 \Big|  -\dfrac{2m}{\frac{\sigma}{m}+1} + K'\,,
\label{uinx}
\end{equation}
where $K'$ is a constant identifying different geodesics. These curves start at past null infinity $\mathscr{I}^-$, diagonally cross the exterior region and end at the future horizon, that is $r\rightarrow m$ and $u\rightarrow +\infty$ for $\sigma \rightarrow (-m)^{-}$.

Naturally, these outgoing and ingoing null geodesics do not abruptly start or end at the past or future horizon, but the null coordinates $u$ and $v$ are not able to follow them past the horizons. However, the leading behavior of \cref{voutx,uinx} near the horizons once again suggest a natural change of coordinates to extend $u$ and $v$ beyond the horizons.

Starting from the leading behavior of \cref{voutx,uinx} near the horizons, and skipping all the intermediate steps detailed in \cref{secglobalKS} for the non-extremal case, a global Kruskal-Szekeres coordinate chart for the maximal extension of the extremal Reissner-Nordström spacetime can be defined by the following transformation:
\begin{widetext}
\begin{equation}
u(U)=
 \begin{cases}
    \quad\quad\quad : &\medskip\\
   {\displaystyle -  \frac{g_{\downarrow}(U,n-1,n)}{U-(n-1)} - \frac{g_{\uparrow}(U,n-1,n)}{U-n}}  
   &\quad\quad n-1< U  <n \, ,\medskip\\ 
      {\displaystyle- \frac{g_{\downarrow}(U,n,n+1)}{U-n} - \frac{g_{\uparrow}(U,n,n+1)}{U-(n+1)}}
      &\quad\quad n< U  < n+1 \, ,\medskip\\
    \quad\quad\quad : &\\
\end{cases}
\label{uUx}
\end{equation}
\begin{equation}
v(V)=
 \begin{cases}
    \quad\quad\quad : &\medskip\\
   {\displaystyle -  \frac{g_{\downarrow}(V,n-1,n)}{V-(n-1)} - \frac{g_{\uparrow}(V,n-1,n)}{V-n}}  
   &\quad\quad n-1< V  <n \, ,\medskip\\ 
      {\displaystyle- \frac{g_{\downarrow}(V,n,n+1)}{V-n} - \frac{g_{\uparrow}(V,n,n+1)}{V-(n+1)}}
      &\quad\quad n< V  < n+1 \, ,\medskip\\
    \quad\quad\quad : &\\
\end{cases}
\label{vVx}
\end{equation}
\end{widetext}

with $n\in\mathbb{Z}$. The position of the horizons in the $(U,V)$ coordinates is arbitrary. I made the choice $U=n$ and $V=n$ for $n\in\mathbb{Z}$.

Defining the function $f(X)$ in such a way that $u'(U)=f(U)$ and $v'(V)=f(V)$, the extremal Reissner-Nordström metric in the Kruskal-Szekeres coordinates $(U,V,\theta,\phi)$ reads
\begin{equation}
\diff s^2 = - h(r) f(U)f(V) \diff U \diff V + r^2 \diff \Omega^2\,,
\label{metricUVx}
\end{equation}
where $r=r(U,V)$ is implicitly defined by \cref{ruvimp,r*x,uUx,vVx}. These equations can then be used to study the regularity of both the function $r(U,V)$ itself and of the metric tensor near the horizons. The analysis is carried out exactly as it is carried out for the non-extremal case, and it shows that the metric tensor in \cref{metricUVx} is well behaved on all spacetime horizons and that the Kruskal-Szekeres null coordinates $(U,V,\theta,\phi)$ provide a smooth and global coordinate chart for the maximal extension of extremal Reissner-Nordström geometry.

The transformation in \cref{uUx,vVx} separately gives a well-defined diffeomorphism between $(U,V,\theta,\phi)$ and $(u,v,\theta,\phi)$ in each spacetime region covered by the latter, even if an analytical expression for the inverse transformation cannot be given. Once again, this is ensured by the strictly monotonicity of $u(U)$ and $v(V)$ in each spacetime region covered by the $(u,v,\theta,\phi)$ coordinate system and by the inverse function theorem. Being the null coordinates $(u,v,\theta,\phi)$ singular on the horizons, also the transformation in \cref{uUx,vVx} is ill defined there.

Kruskal-Szekeres spatial $x_{KS}$ and temporal $t_{KS}$ coordinates can be defined as in \cref{xtKS}. The Kruskal-Szekeres diagram of the maximal extension of extremal Reissner-Nordström spacetime is reported in \cref{figKSdiagramx}.

\section{Conclusions}
\label{secconclusions}

I have derived a smooth Kruskal-Szekeres coordinate chart covering the full maximal extension of the non-extremal Reissner-Nordström geometry. This coordinate chart provides a global generalization to the standard inner and outer Kruskal-Szekeres coordinates and a smooth generalization to the $C^{1}$ global Kruskal-Szekeres coordinate chart derived in~\cite{HamiltonGR}. The same construction has been applied also to the extremal case, obtaining a smooth and global Kruskal-Szekeres coordinate chart for the maximal extension of the extremal Reissner-Nordström geometry.

The existence of this coordinate chart, which is an interesting fact in and of itself, provides a simple alternative to the standard coordinate chart~\cite{Carter:1966zza,CarterRNext,Graves:1960zz,Hawking_Ellis,Chandra} for the conformal diagram of the Reissner-Nordström spacetime. Both of these coordinate charts have the property of bringing ``infinity'' to a finite coordinate value, thus making them a valuable resource to visualize and study the global causal structure of the spacetime. The standard coordinates of the conformal diagram achieve this by performing any generic coordinate transformation that matches the divergence pattern of the different double-null coordinates $(u,v)$ patches and that brings their boundary at a finite coordinate value. This construction leads to a metric which is at most $C^{2}$. By using a transformation that exactly matches the divergence behavior of the null geodesics near the horizons, the global Kruskal-Szekeres coordinates defined here leads to a $C^{\infty}$ metric.

Together with the (Israel-)Klösch-Strobl coordinate chart~\cite{Klosch:1995bw}, they provide the only smooth and global coordinate chart for the maximal extension of Reissner-Nordström geometry. Contrary to the global Kruskal-Szekeres coordinates defined here and the standard coordinates of the conformal diagram, in which the metric contains implicitly defined functions, the metric in (Israel-)Klösch-Strobl coordinates is completely explicit. This makes them more suitable for analytic investigations. These coordinates however do not have the property of bringing ``infinity'' to a finite coordinate value, thus making the diagrammatic representation of global properties less transparent.

Finally, the fact that the global Kruskal-Szekeres coordinates defined here can be straightforwardly generalized to any static spherically-symmetric black hole spacetime whose geometry include multiple horizons makes it an appealing coordinate chart also in different contexts. It was in fact recently used in~\cite{PG_OS} to draw the Kruskal-Szekeres diagram of the geometry of a quantum modification of the Oppenheimer-Snyder model.

\begin{acknowledgments}

I thank Francesca Vidotto and Carlo Rovelli for their comments on an early version of this article. I would also like to thank the anonymous referees for their valuable comments and suggestions, which helped me in improving the quality of the article.

The author's work at Western University is supported by the Natural Science and Engineering Council of Canada (NSERC) through the Discovery Grant ``Loop Quantum Gravity: from Computation to Phenomenology". Western University is located in the traditional lands of Anishinaabek, Haudenosaunee, L\=unaap\`eewak, and Attawandaron peoples.
\end{acknowledgments}

\bibliography{GKS}

\end{document}